\documentclass{webofc}

\usepackage[varg]{txfonts}   
\usepackage{hyperref}
\usepackage{url}
\hypersetup{colorlinks=true,citecolor=blue,urlcolor=blue,linkcolor=blue}
\usepackage{float}
\usepackage{algpseudocode}
\begin{document}
\title{Data Movement Manager (DMM) for the SENSE-Rucio Interoperation Prototype}

\author{
    \firstname{Aashay} \lastname{Arora}\inst{1}\thanks{\email{aashay.arora@cern.ch}} \and
    \firstname{Diego} \lastname{Davila}\inst{1} \and
    \firstname{Jonathan} \lastname{Guiang}\inst{1} \and
    \firstname{Frank} \lastname{W\"urthwein}\inst{1} \and
    \firstname{Harvey} \lastname{Newman}\inst{2} \and
    \firstname{Justas} \lastname{Balcas}\inst{3} \and
    \firstname{Tom} \lastname{Lehman}\inst{3} \and
    \firstname{Xi} \lastname{Yang}\inst{3}
}

\institute{
    University of California San Diego / San Diego Supercomputer Center
    \and
    California Institute of Technology
    \and
    Energy Sciences Network (ESNet)
}

\abstract{
    The Data Movement Manager (DMM) is a prototype interface that connects CERN’s data management software, Rucio, with the Software-Defined Networking (SDN) service SENSE by ESnet. It enables SDN-enabled high-energy physics data flows using the existing worldwide LHC computing grid infrastructure.
    A key feature of DMM is transfer-priority-based bandwidth allocation, optimizing network usage. Additionally, it provides fine-grained monitoring of underperforming flows by leveraging end-to-end data flow monitoring. This is achieved through access to host-level (network interface) throughput metrics and transfer-tool (FTS) data transfer job-level metrics.
    This paper details the design and implementation of DMM.
}
\maketitle
\section{Introduction} \label{sec:intro}

    As we enter the exascale computing era for large collaborative experiments such as the High Luminosity Large Hadron Collider (HL-LHC), the exponential growth of data presents significant challenges for all facets of the scientific computing infrastructure—one of which is networking \cite{zurawski2021hep}.
    Global collaborations rely on efficient data transfer, yet network resources remain limited. The current data transfer model, which relies on re-queuing and repeatedly pushing data until the transfer succeeds, will soon become unsustainable, necessitating greater accountability. One potential solution is to implement controlled data flows for the largest datasets that dominate network usage.
    Traditional networks, characterized by static paths and fixed configurations, lack the flexibility required to optimize large-scale scientific workflows. Software-defined networking (SDN) addresses this limitation by decoupling the control plane (which dictates routing decisions) from the data plane (which handles data movement), enabling dynamic and programmable network configurations. SDN facilitates bandwidth-allocated paths, ensuring that high-priority data flows receive the necessary resources to maximize performance across multiple sites.
    Software Defined Networking for End-to-End Networked Science at the Exascale (SENSE) \cite{monga2020software}, an SDN service developed by ESnet, provides automated and dynamic provisioning of network paths tailored for large-scale scientific data transfers. By optimizing bandwidth usage and reducing latency, SENSE enhances the efficiency of data-intensive workflows. However, configuring a SENSE enabled circuit requires awareness of available network resources and competing user priorities. To streamline this process, we propose integration of SDN capabilities into data management systems like Rucio \cite{Barisits2019-tn}, which inherently manage data transfer priorities \cite{lehman2022data, wurthwein2022managed}.
    The Data Movement Manager (DMM) serves as a prototype interface that bridges Rucio’s data management capabilities with SENSE’s SDN service, enabling SDN-optimized data transfers across the Worldwide LHC Computing Grid (WLCG). This integration allows for the prioritization of critical data transfers, ensuring their timely completion while continuously monitoring network and transfer-level metrics to identify and mitigate underperforming flows. DMM continuously monitors both host-level and transfer-level performance metrics to detect and respond to underperforming flows. Its monitoring infrastructure leverages tools such as Prometheus, Node Exporter, and CERN’s MonIT to correlate real-time throughput with job-level transfer behavior, enabling informed adjustments and diagnostics. Additionally, through its interaction with the File Transfer Service (FTS), DMM dynamically tunes transfer concurrency to saturate available bandwidth efficiently, thereby improving throughput and overall system performance. Architectural design, system interactions, and implementation details of DMM are presented in the following sections.
    DMM is currently deployed in a controlled testbed that closely mimics a production environment, with data transfers carried out over wide-area networks (WAN) between a small number of CMS-affiliated sites. While limited in scale, this deployment enables realistic testing of DMM’s core features and validates its integration with Rucio, SENSE, and FTS. More information about the testbed and the current status of the project is provided in Section \ref{sec:ongoing-future-dev}.

\section{DMM Design Overview} \label{sec:dmm-overview}
    DMM has a daemon-based architecture, which is to say each component runs on a separate thread, ensuring robust and continuous management of data flow requests, (hereafter request). The system maintains the state of each request in a SQL database, which enables persistence and recovery even in the event of a restart. Each request is evaluated by a dedicated daemon based on its current state. Newly added requests start in the INITIALIZED state, where the first step involves allocating SENSE controlled endpoints. Sites supporting SENSE are configured to expose their storage system through multiple endpoints which are necessary to build different isolated SENSE circuits; DMM maintains a list of these endpoints (as detailed in Section \ref{sec:dmm-sense}). Once the endpoints are allocated, the request transitions to the ALLOCATED state. All active requests are then considered for making bandwidth allocation decisions, more details of how this is done are presented in Section \ref{subsec:dmm-core-decision}. After finalizing bandwidth allocations, the SENSE circuit is provisioned and the request advances to the PROVISIONED state. Upon completion, the request enters the FINISHED state. Notably, SENSE circuits associated with finished requests are maintained for a defined period with a minimized bandwidth allocation, allowing for potential reuse by new requests, this is done because provisioning a new circuit is computationally expensive while modifying existing circuits is not; if a circuit remains unused beyond this period, it is subsequently taken down. There are intermediate states between the ones mentioned above which are used for error handling and other internal operations.

\subsection{DMM-Rucio Interaction} \label{sec:dmm-rucio}
    The interactions between DMM and Rucio are bi-directional. From DMM to Rucio, communication is primarily facilitated through the Rucio Python client. DMM periodically queries Rucio to identify newly created transfer rules. Upon detecting a new rule, DMM retrieves its metadata, including the source and destination sites, rule priority, and total transfer size and updates its internal state accordingly.

    Subsequently, Rucio communicates with DMM via DMM’s REST API by providing the rule ID, in response to which DMM returns the corresponding pre-allocated SENSE endpoints. To enable the replacement of transfer endpoints with those managed by SENSE, a small patch is introduced into Rucio which can be summarized as the following pseudocode:

    \begin{center}
        \fbox{
            \begin{minipage}{0.7\linewidth}
                \begin{algorithmic}
                    \If{rule\_needs\_sense()}
                    \For{transfer \textbf{in} transfers\_to\_be\_submitted\_to\_fts}
                    \State new\_endpoints = QUERY\_DMM(transfer.rule\_id)
                    \State transfer.replace(current\_endpoints, new\_endpoints)
                    \State .. .. .. 
                    \State [existing rucio code] 
                    \State .. .. .. 
                    \State transfer.submit\_to\_fts()
                    \EndFor
                    \EndIf
                \end{algorithmic}
            \end{minipage}
        }
    \end{center}

    These changes cause no disruption to the regular non-SENSE Rucio workflows and minimal delay to those leveraging SENSE. The endpoint allocation request is a rather quick operation that is decoupled from the circuit provisioning which happens in the background as rules follow their usual journey.

    It is important to note that SENSE will only be used for point to point transfer rules, that is, rules between a single source and destination site. Given the complexity of multi-source or multi-destination rules, at this time, they will not be considered for SENSE allocation.
    
    Additionally, DMM also continuously monitors each active rule for any modifications in priority, upon detecting such, DMM is able to update the current bandwidth allocations to accommodate the priority change. Once a rule is marked as completed in Rucio, DMM releases the allocated endpoints, making them available for subsequent transfer rules. 
    
\subsection{DMM-SENSE Interaction} \label{sec:dmm-sense}
    DMM interfaces with the SENSE SDN service through a REST API, enabling dynamic provisioning and real-time adjustments of network resources. Initially, DMM retrieves the list of available SENSE endpoints across all configured sites, establishing a comprehensive view of the bandwidth capabilities of each site that supports SENSE. When provisioning a new SENSE circuit, DMM identifies a free endpoint at each of the two sites involved, calculates the bandwidth to be allocated based on relative priorities and optimal usage and requests such allocation to SENSE. In scenarios where the priority of an ongoing data transfer rule is modified, DMM promptly adjusts the circuit's bandwidth to reflect the new priority level. Furthermore, DMM is designed to gracefully decommission circuits when they are no longer required, thereby freeing up network resources for future high-priority transfers.
    
\subsection{DMM-FTS Interaction} \label{sec:dmm-fts}
    The maximum attainable throughput of a network link is directly related to the number of concurrent transfers in flight \cite{Arora2024}. To optimize data flow and fully utilize allocated bandwidth, it is essential to tune the number of active jobs in the File Transfer Service (FTS) \cite{fts} based on both the available bandwidth and the latency between sites. The Data Movement Manager (DMM) accomplishes this by interfacing with FTS via its REST API, dynamically adjusting the number of active transfers to saturate the link efficiently. This integration enables precise control over transfer concurrency, ensuring that the network resources are optimally leveraged for high-throughput data movement.

\section{DMM Core Components} \label{sec:dmm-core}
\subsection{Bandwidth Allocation} \label{subsec:dmm-core-decision}
    In the scenario where DMM is the only entity using SENSE and the primary constraint is the port capacity associated with the active VLAN range, DMM assumes that the primary constraint is the bandwidth capacity of each site (hereafter port capacity). Bandwidth allocation is formulated as a linear programming optimization problem. Initially, a transfer topology graph is constructed with sites as nodes and transfer rules as edges. Given that this graph is a multigraph, it is first converted into a simple graph by merging edges that share the same source and destination, with their priorities summed accordingly. The optimization model is then represented by the (N,E) incidence matrix of the network graph, where N denotes the set of nodes and E represents the set of edges. The objective function is designed to maximize the priority-weighted sum of the bandwidths allocated to the edges, ensuring that higher-priority transfers receive proportionally more bandwidth. This optimization is subject to the constraint that the aggregate bandwidth of the edges incident on any given node must not exceed the node’s port capacity, in addition to the obvious constraint of having non-negative bandwidths, starting at the maximum value of the lower bound to ensure that even the lowest priority transfers have a reasonable allocation. Solving this optimization system yields the optimal bandwidth allocation for each transfer rule, thereby aligning resource distribution with transfer priorities. In the scenario where a feasible solution does not exist, the lower bounds on the bandwidths are relaxed to find a suboptimal solution that satisfies the constraints. To illustrate this, consider the following example transfer rule topology graph in Figure \ref{fig:transfer-graph}:

    \begin{figure}[H]
        \centering
        \includegraphics[width=\linewidth,clip]{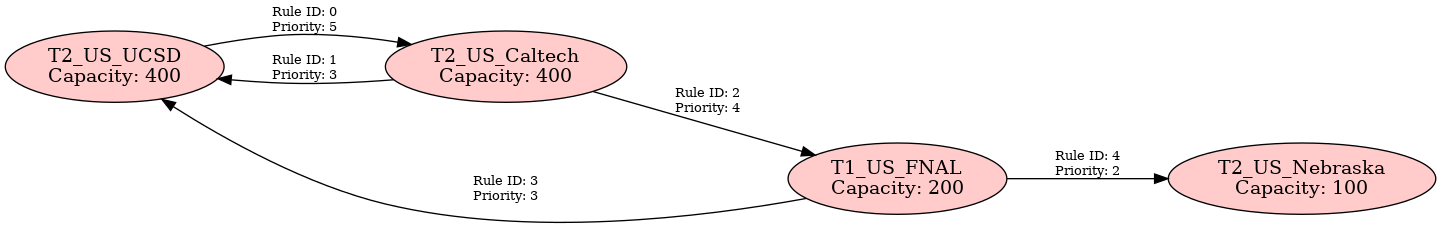}
        \caption{Example Transfer Rule Topology Graph}
        \label{fig:transfer-graph}
    \end{figure}

    The corresponding linear programming optimization problem can be formulated as follows:
    \[
    A =
    \begin{bmatrix}
    1 & 1 & 0 & 0 \\
    1 & 0 & 1 & 0 \\
    0 & 1 & 1 & 1 \\
    0 & 0 & 0 & 1
    \end{bmatrix}
    ,\quad
    b =
    \begin{bmatrix}
    400 \\ 400 \\ 200 \\ 100
    \end{bmatrix}
    ,\quad
    c =
    \begin{bmatrix}
    8 \\ 3 \\ 4 \\ 2
    \end{bmatrix}
    \] 
    $$\textbf{Objective:} \quad \text{Maximizing} \  c^T x $$
    $$\textbf{Subject to:} \quad Ax \leq b, \ x \geq l \geq 0$$

    Where \( A \) is the incidence matrix, \( b \) is the port capacity vector, and \( c \) is the priority vector after merging the edges between each pair of sites and $l$ is the chosen lower bound imposed on the solution. Notice that the graph is directed, but this fact is not relevant for the optimization problem since the bandwidths are assigned to the edges, not the nodes.

    Solving the optimization system yields multiple solutions depending on the chosen lower bound, and the solution with the highest objective value is selected as optimal for the transfer rule topology. In this instance, the optimal solution is 
    \[
    x = [335,\, 65,\, 65,\, 70].
    \]
    The final bandwidth allocation is further refined by distributing the bandwidth for each pair of sites in the original multigraph according to the relative priorities of the corresponding transfer rules. For example, the 335 units allocated for transfers between \texttt{T2\_US\_UCSD} and \texttt{T2\_US\_Caltech} are divided between two rules in proportion to their priorities. The resulting allocations are as follows:
    \begin{small}
        \begin{itemize}
            \item \texttt{T2\_US\_UCSD} $\leftrightarrow$ \texttt{T2\_US\_Caltech}: priority 5, bandwidth 210
            \item \texttt{T2\_US\_UCSD} $\leftrightarrow$ \texttt{T2\_US\_Caltech}: priority 3, bandwidth 125
            \item \texttt{T2\_US\_UCSD} $\leftrightarrow$ \texttt{T1\_US\_FNAL}: priority 3, bandwidth 65
            \item \texttt{T2\_US\_Caltech} $\leftrightarrow$ \texttt{T1\_US\_FNAL}: priority 4, bandwidth 65
            \item \texttt{T1\_US\_FNAL} $\leftrightarrow$ \texttt{T2\_US\_Nebraska}: priority 2, bandwidth 70
        \end{itemize}
    \end{small}

\subsection{Online and Offline Monitoring} \label{subsec:dmm-core-monit}
    SENSE employs monitoring tools such as Prometheus, Node Exporter, and Grafana to facilitate real-time, hop-by-hop monitoring and debugging across multiple domains, thereby ensuring visibility into network path performance. In addition, DMM leverages job-level metrics obtained from CERN’s MonIT infrastructure to access detailed information on FTS transfers, including failures, retries, and overall performance. Through periodic queries to Prometheus, DMM continuously monitors the throughput of ongoing transfers and can dynamically adjust bandwidth allocations in response to real-time performance data. By correlating these host-level metrics with job-level insights from FTS, DMM generates comprehensive reports for each transfer, providing deep visibility into system performance and identifying potential issues for prompt remediation.

\subsection{Frontend} \label{subsec:dmm-core-frontend}
    DMM features a user-friendly frontend dashboard that displays all active requests managed by the system, along with the current status of each site. This interface provides a comprehensive overview of ongoing data transfers, enabling users to monitor the progress of their requests and track the performance of individual sites in real time.

\section{Ongoing and Future Development} \label{sec:ongoing-future-dev}
    DMM is currently deployed within a controlled testbed environment involving a limited number of CMS-affiliated sites \cite{Davila_2024}. It manages approximately ten concurrent data transfer requests and operates over high-capacity wide-area network (WAN) links ranging from 100 to 400 Gbps. These transfers are conducted using dedicated data transfer nodes (DTNs) configured with storage systems that are separate from those used in production workflows, ensuring that testing remains isolated while still representative of real-world conditions.

    The next phase of development focuses on transitioning DMM into a pseudo-production environment, beginning with its integration into CMS’s integration instance of Rucio. This step will enable broader testing under more realistic operational constraints. Furthermore,  while DMM currently functions as a standalone service, there is considerable potential to evolve it into a fully integrated component of Rucio. Such integration is expected to streamline workflow management, reduce operational overhead, and enhance the scalability and robustness of the system as it moves toward wider deployment.

\section{Conclusion} \label{sec:conclusion}
    The integration of SDN services like SENSE with data management systems such as Rucio represents a significant advancement in optimizing large-scale scientific data transfers. By dynamically provisioning network paths, adjusting bandwidth allocations in real time, and fine-tuning active transfers based on link characteristics, the system would ensure efficient use of network resources in the exascale era. Additionally, comprehensive monitoring through tools like Prometheus, Node Exporter, and FTS metrics could provide critical insights that facilitate proactive management and troubleshooting of data flows. 
    Ongoing and future developments, including the transition to pseudo-production environments and deeper integration with Rucio, promise to further streamline operations and enhance performance. Collectively, these innovations offer a robust framework for managing the increasingly complex demands of global scientific collaborations, paving the way for improved throughput, reliability, and overall efficiency in high-performance and high-throughput computing environments.

\section{Acknowledgements} \label{sec:acknowledgements}
    This ongoing work is partially supported by the US National Science Foundation (NSF) Grants OAC-1841530, OAC-1836650, PHY-2323298 and PHY-1624356.
    In addition, the development of SENSE is supported by the US Department of Energy (DOE) Grants DE-SC0015527, DE-SC0015528, DE-SC0016585, and FP-00002494. 
    Finally, this work would not be possible without the significant contributions of collaborators at CENIC, ESnet, Caltech, and SDSC.

\bibliography{references.bib}

\end{document}